\documentstyle[preprint,aps]{revtex}
\tighten
\begin{document}
\title{Constraints on Black Holes in Classical and Semiclassical Theories
        of Gravity}
\author{Paul R. Anderson\cite{PRA} and Courtney D. Mull\cite{CDM}}
\address{Department of Physics\\Wake Forest University\\Winston-Salem, NC  27109}
\maketitle
\begin{abstract}
Constraints on the geometry of a static spherically symmetric black hole are obtained by requiring the spacetime curvature to be analytic at the event horizon.  For a zero temperature black hole further constraints are obtained by also requiring that the semiclassical trace equation be satisfied when conformally invariant fields are present.  It is found that zero temperature black holes whose sizes lie within a certain range do not exist.  The range depends on the numbers and types of conformally invariant quantized fields that are present.

\end{abstract}
\section{Introduction}

One of the unanswered questions in semiclassical gravity is how quantized fields alter the spacetime geometry near the event horizon of a black hole.  This is important because the thermodynamic properties of a black hole are determined by the geometry at the event horizon.  Some work has been done to answer this question using linearized semiclassical gravity and either analytical approximations or numerical computations of the stress-energy tensor in Schwarzschild spacetime\cite{York,HK,HKY,AHWY,Winnipeg}.  Further progress has been hampered by problems with the analytical approximations near the event horizons of other black holes\cite{FZ,AHS,AHL} and by the difficulty involved in numerically computing the stress-energy tensor for quantized fields in black hole spacetimes.  

Given these difficulties and the importance which quantum effects may have in black hole spacetimes, it is useful to see if constraints can be placed on the possible geometry's that black holes can have at their event horizons.  In this paper constraints relating to static spherically symmetric black holes are found using two methods.  The first is very general and applies to both classical and semiclassical metric theories of gravity.  It is to assume that the spacetime curvature is analytic at the
event horizon.  This assures that there will be no curvature singularities there.
This is a very reasonable assumption that is satisfied by all known vacuum
solutions to Einstein's equations that correspond to black holes.  Further it
is not without precedent.  In one of the uniqueness theorems which Hawking\cite{Hawking} has proven for rotating stationary black holes it is assumed that the metric is analytic at the event horizon.  Recently Zaslavskii\cite{Z} has determined the general form of the metric 
near the event horizon of near extreme and extreme charged black holes when the black holes are 
in a cavity and the grand canonical ensemble is utilized.  
He assumed that the metric can be expanded in a power series in the radial coordinate
$r$ near the event horizon.

The requirement that the curvature be analytic at the event horizon of a static black hole results
in significant constraints on the geometry in that region.
These are greatest for nonzero temperature black
holes where it is found that the spacetime geometry must be of the same general form near
the event horizon as that of Schwarzschild and Reissner-Nordstr\"{o}m black holes.  

The second method involves using the trace equation in semiclassical gravity to constrain
the spacetime geometry near the event horizon of a zero temperature black hole. 
In general it is extremely difficult to solve the semiclassical back reaction equations because of the difficulty in computing the stress-energy tensor for quantized fields in a spacetime with an arbitrary or at least somewhat arbitrary geometry.  However, for conformally invariant fields it is well known that the trace of the stress-energy tensor is exactly equal to the trace anomaly which is known analytically in a general spacetime.  Since any solution to the semiclassical equations must be a solution to the trace equation, one can put constraints on solutions to the semiclassical equations by examining the trace equation.  In general such constraints are not very interesting because they are not very strong.  
However if the curvature is analytic at the event horizon and if only conformally invariant
fields are present, then it is found that there is a range of sizes for which zero
temperature black holes cannot exist.  The range of excluded sizes depends on the 
number and types of quantized fields present.  In many cases this range extends 
to zero, resulting in a minimum size for zero temperature black holes.   

In what follows we first discuss restrictions which result from requiring the curvature to be analytic at the event horizon.  Then those resulting from the trace equation are given.

\section{Constraints due to the curvature}

The metric for a static spherically symmetric spacetime can be written in the general form\cite{units}
\begin{equation}
d s^2 = - f(r) dt^2 + \frac{1}{k(r)} dr^2 + r^2 d \Omega^2 \;\;\;.
\end{equation}
The unique nonvanishing components of the Riemann curvature tensor in an orthonormal frame are
\begin{mathletters}
\begin{eqnarray}
R_{\hat{t} \hat{r} \hat{t} \hat{r}} &=&  \frac{v' k}{2} + \frac{v k'}{4} + \frac{v^2 k}{4} \\
R_{\hat{t} \hat{\theta} \hat{t} \hat{\theta}} &=&  \frac{v k}{2 r} \\
R_{\hat{r} \hat{\theta} \hat{r} \hat{\theta}} &=& - \frac{k'}{2 r} \\
R_{\hat{\theta} \hat{\phi} \hat{\theta} \hat{\phi}} &=& \frac{1 - k}{r^2}\;\;,
\end{eqnarray}
\end{mathletters}%
where $v \equiv f'/f$ and primes denote derivatives with respect to $r$.
If the spacetime has an event horizon then $f$ vanishes on that horizon.    
The surface gravity at the event horizon is given by the formula 
\begin{equation}
\kappa = \frac{v}{2} \left(f k \right)^{1/2} \;\;.
\end{equation}

We require that the above components of the Riemann tensor be analytic at the event 
horizon.  From Eq.(2d) it is clear that $k$ must then be analytic at the event horizon. 
Eq.(2b) shows that the quantity $v\, k$ must also be analytic.  This second condition 
results in the further condition that $k \rightarrow 0$ at the event horizon.  
To see this note that $v$ must diverge at the horizon because $v = f'/f$ and $f$ 
vanishes there.  These conditions can be summarized by saying that near the event 
horizon $v$ and $k$ have the following leading order behaviors:
\begin{mathletters}
\begin{eqnarray}
v &=& p (r-r_0)^{m-n} \\
k &=& q (r-r_0)^n  \;\;\;.
\end{eqnarray}
\end{mathletters}%
Here $p$ and $q$ are real positive constants, $m$ and $n$ are integers which satisfy the condition $n > m \ge 0$, and $r_0$ is the value of $r$ at the event horizon.

Further restrictions can be obtained by considering Eq.(2a).  Substituting Eqs.(4a-b) 
into (2a) one finds that
\begin{equation}
R_{\hat{t} \hat{r} \hat{t} \hat{r}} = -\frac{p\, q}{4}\, (n - 2 m)\,(r-r_0)^{m-1} 
   + \frac{p^2 q}{4}\, (r - r_0)^{2 m - n} \;.
\end{equation} 
These terms on the right hand side of (5) must either be separately finite at the horizon or they must cancel.  They only cancel if $m = n-1$ and $p = 2 - n$.  However we previously showed that $n \ge 1$.  Thus, since $p > 0$, the only case in which they cancel is $n = 1$,\, $m = 0$, \, $p = 1$.  In this case it is easy to see that near the horizon $f = c\, (r - r_0)$ for some positive constant $c$.  If the terms on the right hand side of Eq.(5) vanish separately then the restrictions are $m \ge 1$ and $2 m \ge n$.  

To summarize, either solutions exist with
\begin{mathletters}
\begin{equation}
n = 1, \;\; m = 0, \;\; p = 1, 
\end{equation}
or there is the constraint
\begin{equation}
2m \ge n > m \ge 1
\end{equation}
\end{mathletters}%
In the first case the surface gravity and hence the temperature is finite and nonzero.  In the second case it is always zero.  Thus all static spherically symmetric black holes with nonzero temperatures must have 
\begin{eqnarray}
f &=& c (r - r_0) + ... \nonumber \\
k &=& q (r - r_0) + ...
\end{eqnarray}
near their event horizons if the components of the Riemann tensor in an orthonormal frame are analytic at the event horizon.

\section{Constraints due to the trace anomaly}

Semiclassical gravity can be used to place further constraints on the geometry near the event horizon of a static spherically symmetric black hole in the case that only conformally invariant free quantized fields are present.  The semiclassical back reaction equations can be written in the general form
\begin{equation}
G_{\mu\nu} + a A_{\mu\nu} + b B_{\mu\nu} = 8 \pi (T_{\mu\nu})_{cl} + 8 \pi <T_{\mu\nu}>
\end{equation}
Here $A_{\mu\nu}$ and $B_{\mu\nu}$ are tensors which result from the variation of a scalar curvature squared term and a Weyl squared term in the gravitational Lagrangian.  Their coefficients, $a$ and $b$, are arbitrary and must in principle be determined by experiment or observation.  $(T_{\mu\nu})_{cl}$ is the stress-energy tensor for any classical fields.  We shall only be concerned here with the classical electromagnetic field\cite{note1}.  The trace of  $B_{\mu\nu}$ is identically zero as is the trace of the stress-energy tensor for the classical electromagnetic field.  The trace of $A_{\mu\nu}$ is equal to $-6 \Box R$\cite{BD}.  For conformally invariant fields the trace of $<T_{\mu\nu}>$ is equal to the trace anomaly\cite{BD}.  Thus the trace equation is
\begin{equation}
- R - 6 a \Box R = 8 \pi [\alpha \Box R + \beta (R_{\alpha\beta} R^{\alpha \beta} - \frac{1}{3} R^2) + \gamma   
       C_{\alpha\beta\gamma\delta} C^{\alpha\beta\gamma\delta}]
\end{equation}
with
\begin{mathletters}
\begin{eqnarray}
\alpha = [N(0) + 6 N(1/2) - 18 N(1)]/2880 \pi^2 \\
\beta = [N(0) + 11 N(1/2) + 62 N(1)]/2880 \pi^2 \\
\gamma = [N(0) + \frac{7}{2} N(1/2) - 13 N(1)]/2880 \pi^2 \;\;.
\end{eqnarray}
\end{mathletters}%
Here $N(0)$, $N(1/2)$, and $N(1)$ are the number of scalar, four component spin $1/2$, and vector fields respectively. 

If Eqs.(4a-b) are substituted into Eq.(9) and condition (6b) is imposed, then the 
leading order terms near the horizon can be computed for various values of $m$ and
$n$.  Consider first the case $2 m > n > m \ge 2$.  In the limit $r \rightarrow r_0$ 
Eq.(9) becomes
\begin{equation}
- \frac{16 \pi}{3 {r_0}^2}\, (\beta + 2 \gamma) - \frac{2}{{r_0}^2} = 0
\end{equation}
     From Eqs.(10a-c) it is clear that for all fields $\beta + 2 \gamma > 0$.  Thus there
are no solutions to the trace equation for values of $m$ and $n$ in this range.

The only other possibility is $2m = n \ge 2$.  This means that the only possible 
zero temperature solutions to the semiclassical back reaction equations have the form
\begin{eqnarray}
v &=& p (r - r_0)^{-m}  \nonumber \\
k &=& q (r - r_0)^{2 m}  \;\;\;,
\end{eqnarray}
near the event horizon, with $m \ge 1$.
Substituting (12) into (9) and taking the limit $r \rightarrow r_0$ results in the
equation
\begin{equation}
\frac{p^2 q}{2} - \frac{\pi \beta p^4 q^2}{3} - \frac{2 \pi \gamma p^4 q^2}{3} - \frac{16 \pi (\beta + 2 \gamma)}{3 {r_0}^4} - \frac{2}{{r_0}^2} 
- \frac{16 \pi (\beta - \gamma) p^2 q}{3 {r_0}^2} = 0 \;\;.
\end{equation}
This equation has the solutions
\begin{equation}
q_{\pm} = \frac{1}{4 \pi (\beta + 2 \gamma) p^2 {r_0}^2} \, [3 {r_0}^2 - 32 \pi (\beta - \gamma) \pm (768 \pi^2 \beta^2 - 3072 \pi^2 \beta \gamma - 288 \pi \beta {r_0}^2 + 9 {r_0}^4 )^{1/2}] \;.
\end{equation}
It is interesting to look these solutions in the limit $r_0 \rightarrow \infty$.  They are
\begin{mathletters}
\begin{eqnarray}
q_+ &\rightarrow& \frac{3}{2 \pi (\beta + 2 \gamma) p^2} \\
q_- &\rightarrow& \frac{4}{p^2 {r_0}^2} 
\end{eqnarray}
\end{mathletters}%
The second solution has the same type of behavior as the extreme Reissner-Nordstr\"{o}m solution (for which p=2).  The first has a different behavior. When substituted into Eqs.(2a-d) 
it is seen that the spacetime curvature at the event horizon is large for this solution even
in the limit $r_0 \rightarrow \infty$.  Therefore this is not a physically acceptable solution for large $r_0$ since quantum effects would be large even on macroscopic scales.  It should be
emphasized here that there are many solutions to the trace equation which are not solutions
to the full set of semiclassical back reaction equations.
 
Finally it is possible to put bounds on the allowed values of $r_0$ for zero temperature black
holes.  Since the 
left hand side of (14) is positive and real, the right hand side must be also.  However the right hand side is complex if
\begin{eqnarray}
r_{-} &<& r_0 <r_{+} \nonumber \\
 r_{\pm} &=& 4 (\pi \beta)^{1/2} \left[1 \pm \left(\frac{2}{3 \beta} \right)^{1/2} (\beta + 2 \gamma)^{1/2} \right]^{1/2} \;\;\;.
\end{eqnarray}
For all allowed values of $\beta$ and $\gamma$\, $r_{+}$ is real.  If 
$\beta < 4 \gamma$ then $r_{-}$ is imaginary and solutions only occur for  
$r_0 > r_{+}$.  If $\beta > 4 \gamma$ then solutions also occur for $0 < r_0 < r_{-}$.

	We have examined the constraints on static spherically 
symmetric black holes which are imposed by requiring that the components of the 
Riemann tensor in an orthonormal frame be analytic at the event horizon.  This 
requirement forces the metric functions for nonzero temperature black holes to be of 
the form (7).  For zero temperature black holes we further imposed the condition 
that the semiclassical trace equation be satisfied in the case that only conformally invariant fields are present.  This resulted in the constraint that zero temperature black holes 
must have metric functions 
of the form (12).  Also no zero temperature black holes can exist in this case
with values of $r$ at the event horizon in the range $r_{-}< r_0 < r_{+}$.
Here $r_{\pm}$ is given by Eq.(16) with the understanding that if $r_-$ is complex, the
lower limit is zero.

\acknowledgments

	This work was supported in part by Grant No. PHY95-12686 from the National Science Foundation.


\begin{references}
\bibitem{note1} In fact the results will be valid for any classical stress-energy tensor whose trace vanishes.
\end{references}
\end{document}